# Boost the Public Demand for Soft Matter Education and Career Opportunities with a Homemade Video

*An Interactive Qualifying Project*

*Submitted to the Faculty*

*of*

*WORCESTER POLYTECHNIC INSTITUTE*

*In partial fulfillment of the requirements for the*

*Degree of Bachelor of Science*

by

Yen-Chen Chen

Kun-Ta Wu PhD, Advisor

# Table of Contents

# Abstract

This study aims to promote the awareness and education of soft matter physics which has recently become a popular research topic due to its capability of developing self-assembling materials for numerous industries such as medical treatments and surgeries. First, a homemade video is designed to illustrate the concepts of soft matter to general public and students. Next, an online survey is conducted which assesses people's interest in soft matter before and after watching the video, whether more government investments on soft matter education are needed, and the distribution of related resources. Statistics indicate that the video has effectively stimulated the overall interest of soft matter knowledge, research, and careers by the average of 27.8 %. Within different student majors, after watching the video, chemical and biomedical engineering students interested in soft matter the most whereas physical engineering students have the largest growth in soft matter interest. Within different school years, the third- and fourth-year undergraduate students have the largest improvement on their understanding to soft matter. Within different degrees students plan to pursue, PhD students interested in soft matter the most whereas master or bachelor students have the largest growth in soft matter interest. Furthermore, students with mid-level interest in physics, biophysics, and material science, and high level of interest in biochemistry and mathematics have higher improvement on their interest and understanding to soft matter. Finally, most respondents agree on the increment of government funding of soft matter research and the offer of related college courses, and the courses should be offered in advanced level within the departments of physics, biomedical engineering, and chemistry and biochemistry. In conclusion, our video has stimulated people's awareness on soft matter so that it becomes a consensus to have more investments on soft matter education.

# Authorship

The project was implemented by Yen-Chen Chen, including producing the soft matter video, designing survey questions, acquiring and analyzing data, statistics, and writing the report.

# Acknowledgement

I would like to acknowledge, the Department of Physics and Department of Mechanical Engineering at Worcester Polytechnic Institute and the students who watched my video about soft matter and answered the online survey on soft matter education. Furthermore, I would also like to thank my advisor Kun-Ta Wu, Ph.D. for his supervision and encouragement during this project.

# Chapter 1: Introduction

Physics is everywhere in our daily life. From fire that boils the water to the electricity that powers the city, human civilization cannot be established without the deep understanding to the principles and applications of physics. Therefore, in the past five centuries, people have been diligently developing new theories to explain the existing phenomena as well as performing research to discover the things unknown. For example, Isaac Newton introduced the three Newton's Law of Motion in 1750 to describe the relationship between a macroscopic system and the forces acting upon it, and its motion in response to those forces [7]. After the classical macroscopic physics have been deeply understood, people started to focus on understanding the physical phenomena at microscopic scale, such as atom, molecule, and quantum. The most significant contributor to the knowledge of microscopic physics was Erwin Schrödinger who developed the Schrödinger equation, a set of linear partial differential equation that describes the wave function or state function of a microscopic quantum-mechanical system, in the early twentieth century [8]. As the microscopic physics have been understood to a certain extent, physicists started to be curious about the unknown physical phenomena occurred at the scale between the traditional macroscopic and microscopic scales. In 1981, Van Kampen introduced the concept of "mesoscopic scale" which is the scale between $10^{-9}$ to $10^{-6}$ meter and is where both the quantum physics and classical mechanics meet [6] [24]. At this scale, one of the important subfields called "soft matter physics [9]," which include the physical properties related to liquids, colloids, polymers, foams, gels, granular materials, liquid crystals, and most importantly, the biological materials, is receiving growing attention since 2011 when the US Department of Energy (DOE) started discussing and exploring the potential of mesoscale science and technology at its Basic Energy Sciences Advisory Committee [21]. More and more physicists started to concentrate on soft matter research because one of its most important physical properties, that is molecular self-assembly powered by diffusion, enables us to develop smart materials that can be programmed to grow by themselves. Furthermore, active matter [1], one of the subdisciplines of soft matter which consists of molecular motors that can convert chemical fuels into kinetic energy and propels the biological system, has become increasingly important because not only it provides an alternative to power the mesoscopic machine but it also pushes people closer to the creation of artificial life by understanding how biological system transport materials in cells.

While soft matter and active matter have demonstrated their rich academic value, they also lead to miscellaneous industrial applications and thus career opportunities. The most obvious example is that oil industries are actively hiring soft matter scientists with high salary (ranging from $100K to $130K per year according to Glassdoor [3]) to conduct research on improving the oil refinery process, such as extracting oil from water-oil mixture or understanding how liquid goes through porous materials. The background in soft matter also allows people to work in cosmetic companies and cleaning companies because cosmetics and shampoos are both part of soft matter and developing these products requires a background in soft matter research. Soft matter also plays an important role in medical health. For example, developing a microfluidic device that can efficiently select active "sperms" has become a fast-growing business to solve the problem of infertility. Designing this device involves the knowledge of controlling the fluid at "mesoscopic" scale, thus within the scope of soft matter. Despite soft matter related research and products play an increasingly critical role in both academia and industries within the past decade, only 307 undergraduate degrees were awarded in biophysics in the year of 2017 [4] and most of them were from universities that were ranked the top 100 best national universities by US News and World Report in the same year [10]. This shows that the educational resources on soft matter physics and biophysics for undergraduate students are considerably inadequate around the United States. Therefore, the aim of this study is to promote the knowledge and education on soft matter related physics phenomena at Worcester Polytechnic Institute (WPI) through a homemade video. In addition, we will conduct an online survey for assessing the student understanding as well as awareness of soft matter related research and career opportunities before and after watching our video. The statistics of the survey results will then be used to evaluate whether WPI should offer a biophysics course for undergraduates and how to distribute the related educational resources at different departments as well as student school years.

# Chapter 2: Literature Review

Starting from the early 2000s, many research works have been conducted to examine the physical properties of soft matter and active matter as well as their applications to create programmable self-assembling materials or other potential use in medical treatments. In this chapter, we will briefly introduce and summarize the concepts and recent research progress on soft matter and active matter. In addition, we will quickly review all the existing studies and surveys regarding to the biophysics education for undergraduates.

## 2.1: Soft Matter

Soft matter is a kind of condensed matter consisting various physical systems that can be deformed or structurally altered by thermal or mechanical stress at the magnitude of thermal fluctuations. They include liquids, colloids, polymers, foams, gels, granular materials, liquid crystals, and biological materials such as bacteria and microtubule [9]. Soft matter comprises of mesoscopic constituents with the size between $10^{-9}$ to $10^{-6}$ meter, which are larger than microscopic materials such as atom and molecule but also smaller than macroscopic materials which are visible by naked eyes [6]. At this scale, the motion of the constituents when suspended in water is dominated by Brownian motion which the water molecules collide with the mesoscopic soft matter constituents from all directions, causing the material to explore the space and undergo the process of diffusion [5]. Diffusion is one of the most fundamental mechanism that cause the soft matter to self-assemble. For example, when the colloids are coated with complementary single-stranded DNA, these DNA strands hybridize into DNA bonds that link pairs of colloids together [18] [29]. When water molecules collide with these colloid pairs from all directions, it will cause an unbalanced pressure on the particle surface that driven the particles to meet other colloid pairs in the space and assemble. In a macroscopic scale, these DNA-coated colloids have been reported to self-assemble into crystals with programmable structures [13] [19] [20] [32]. Numerical simulations showing that by designing the interactions between nanoparticles that are directionally functionalized with DNA, one can direct the colloids to slowly self-assemble into a mesoscopic Empire State Building [16]. In addition, the newest research indicates that the self-assembly of DNA-functionalized colloidal particles can be programmed using linkers dispersed in solution which allows hundreds of different interactions needed for assembling the desired mesoscopic structure to happen simultaneously [20]. Research also indicates that the process of self-assembly can be further accelerated by exposing the DNA coated colloidal surfers under UV light within the hydrogen peroxide solution. Under this condition, the colloidal surfers will consume hydrogen peroxide so that a concentration gradient is generated within the solution which will propel the particles and thus accelerate the self-assembly process [23]. One of the main applications to programmable self-assembly is to create a mesoscopic robot [28] that can be used for performing minimally invasive surgery to fight against diseases such as performing treatments for clearing the clogs in the blood artery.

## 2.2: Active Matter

Active matter is another discipline of soft matter which is differentiated from the conventional passive matter by its ability of converting local chemical fuels into kinetic energy [1]. For example, animals eat foods to maintain the living; bacteria consume sugar and oxygen to swim; colloidal surfers consume hydrogen peroxide to self-propel. These active matter constituents interact to collectively form a macroscopic material flow. For example, microtubule-based active fluid comprises of microtubules. These microtubules are bridged by kinesin motor clusters to slide apart. These sliding dynamics accumulate from hundreds of microtubules to form a microtubule bundles that self-extend [12][25]. The bundle extension stirs the surrounding fluid, generating flows. These flows formed random vortices, canceling each other out in a long term. However, research has reported that the motion of active fluid is sensitive to the physical boundary conditions. By confining this active fluid in a circular confinement such as cylinder or toroid can regulate these flows to self-organize from complete turbulent into a coherent flow that circulates around the container, capable of transporting materials in a distance beyond the limit of passive fluid [30]. In addition, it was demonstrated that the bacteria-based active fluid can be used to turn the mesoscopic gears [26]. These results demonstrate a potential of using the active fluid to perform mechanical work in mesoscopic system which serves as an alternative to power generation. However, before people can use active fluid to power the mesoscopic machines, the mechanisms that can control the active fluid flow speed so that the machine can be turned on or turned off according to user settings are indispensable. It was reported that active fluid is sensitive to temperature; by controlling the surrounding temperature, one can speed up and speed down the flow of active fluid instantaneously [11] [12]. Furthermore, researchers have modified the composition of kinesin motor so that its structure is sensitive to light. Thus, one can initiate the flow of active fluid by exposing the system under UV light; conversely, without light exposure, the active fluid flow will automatically stop [27].

In addition to experimental studies, theorists are also working diligently to characterize the motion of active fluid as well as examining the factors that cause the self-organization of active fluid. The most well-known way of doing so is through the simulation of microtubule bundle structure and nematic structure [17] [31]. Recently, the simulation of active fluid motion has become increasingly important as people are trying to commercialize the active-fluid-related products. For example, the simulation allows people to develop a microfluidic device that can select the most active sperm to reduce the infertility rate [14]. In addition, by understanding the physical behaviors of active matter both experimentally and numerically, we can get a glimpse of how eukaryotic cells transport molecules through the fluidic cytoplasm powered by kinesin, which pushes us closer to the ultimate goal of creating artificial cells [22].

## 2.3: Studies and Survey on Biophysics Education and Career Opportunities

In the past decade, education on biophysics has attracted national attention; more and more universities around the country have provided the courses and degree programs in biophysics. However, biophysics is multidisciplinary, in order to let student learn well on biophysics, one must have solid background on mathematics, mesoscopic physics, biology, and fundamental knowledge on biochemistry, which makes the educators difficult to design an educational program on biophysics for students with different background knowledge. Therefore, researchers have developed an online module for undergraduate students that constructs the concepts of biophysics step by step, from molecular structure determination, diffusion, to molecular machines, which is proven to work for students at the University of California, Santa Cruz [15]. In addition, the same research also shows that computer programming is an important skill on biophysics as more researchers are dedicated to performing numerical simulation on biological systems. This demonstrates that biophysical research and education can attract people from disciplines beyond physics and biology. After students receive degree on biophysics, there is no need to worry about career opportunities. Statistics have predicted that the job opportunities related to biophysical research will grow 6 % annually in the next 10 years [2]. Despite the fast-growing market demand on biophysicists, statistics also showed that only 307 undergraduate degrees were awarded in biophysics in the year of 2017 [4] and most of them were from universities that were ranked the top 100 best national universities by US News and World Report in the same year [10], which is considerably rare compares to tens of thousands degrees awarded annually in engineering disciplines. This clearly indicates the extreme shortage of educational resources on biophysics for undergraduates nationwide. Thus, it becomes biophysical researchers' responsibility to promote the related educational programs in their own institutes.

# Chapter 3: Materials and Methods

In this study, we will promote the education of biophysics especially the knowledge related to soft matter and active matter by three processes. First, we will make a short video that briefly introduces the concepts of soft matter and active matter in common words which are understandable to the general public without any prior background knowledge. Secondly, we will design an online survey that assesses the student awareness, understanding as well as personal interest in soft matter physics and related research or career opportunities before and after watching our short video. Finally, after receiving the survey feedbacks, we will then do the statistics on whether or not WPI should offer biophysics courses as well as how to distribute the related laboratorial resources at different departments and students at different school years. The detailed processes of this study will be illustrated in the following sections.

## 3.1: Design of Short Video

In this study, we used the Adobe After Effect (*Figure 3.1.1*) to make the video presenting the concepts of soft matter and active matter which lasts for roughly 7.5 minutes. Below I will briefly illustrate the contents of the video as well as their functions on promoting soft matter educations and careers.

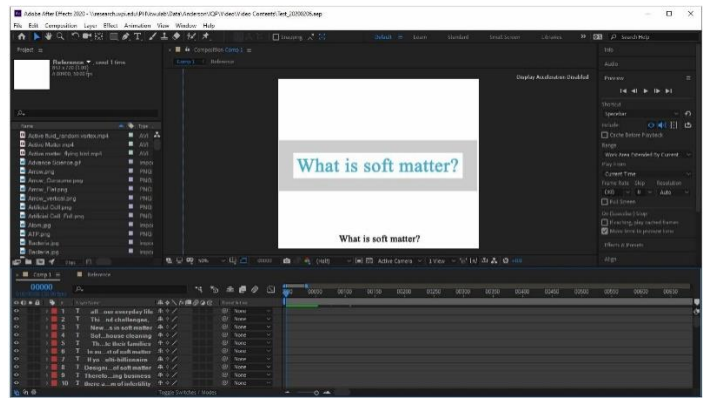

**Figure 3.1.1**: The interface of Adobe After Effect

First, we briefly introduce what is soft matter, including its discipline in physics, major physical properties including the deformable structure by thermal or mechanical stress, and what are the substances included the field of soft matter (***Figure 3.1.2***). We also emphasize the biological materials such as bacteria and microtubule because they will be the focus in the later part of the video. Therefore, the following contents are included in the first paragraph of the video:

*"What is soft matter?*

*Soft matter is a kind of condensed matter consisting of a variety of physical systems that can be deformed or structurally altered by thermal or mechanical stress at the magnitude of thermal fluctuation. They include liquids, colloids, polymers, foams, gels, granular materials, liquid crystals, and biological materials such as bacteria and microtubule."*

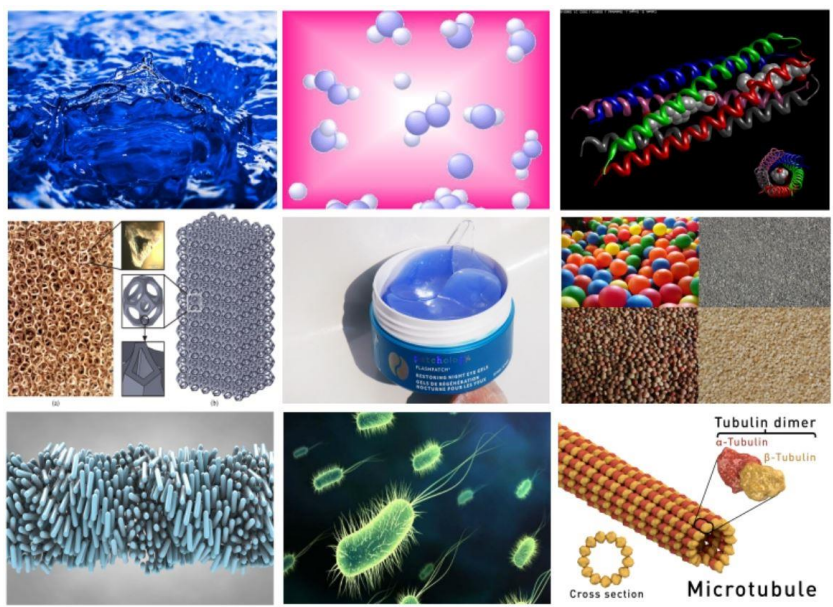

**Figure 3.1.2**: Examples of soft matter materials (extracted from 00:42 of the video)

At the beginning of the second paragraph, we first introduce the size of the soft matter that is mainly mesoscopic (between $10^{-9}$ to $10^{-6}$ meter; ***Figure 3.1.3***), which is between microscopic scale such as quantum, atom, and molecules, and macroscopic scale such as all the substances visible by naked eyes. Next, we introduce the main physical phenomena that dominate the substances within the mesoscopic scale, which is Brownian motion that causes the constituents in the water to move randomly, explore the space, and thus undergoes the process of diffusion (***Figure 3.1.4***). Diffusion is critical for soft matter because it is the fundamental mechanism that causes the soft matter to self-assemble. After introducing the basic physics of soft matter, we then introduce a real-life example and application of soft matter self-assembly, which is colloids coated with complementary single-stranded DNA. In such system, it was reported that DNA strands would hybridize into DNA bonds that link pairs of colloids together which self-assemble into crystals with programmable structure (***Figure 3.1.5***). Here, the programmable structure of self-assembled colloids is considerably important because without this knowledge, scientists cannot let colloids to self-assemble into a useful tool or product in mesoscopic scale. To impress the audiences on the importance of programmable self-assembly, we then introduce a cool numerical simulation showing that if the interactions between each pairs of colloids are well-controlled, one can let the colloids to self-assemble into mesoscopic Empire State Building (***Figure 3.1.6***). Lastly, we briefly mention the potential applications of programmable self-assembly, which is to create the parts for micro-robots to fight against diseases such as performing treatments for clearing the clogs in the blood artery (***Figure 3.1.7***). Therefore, the following contents are included in the second paragraph of the video:

*"Soft matter comprises of mesoscopic constituents which are larger than atoms and molecules but also smaller than macroscopic materials such as a pencil. At this scale, the motion of the constituents when suspended in water is dominated by Brownian motion because the water molecules kick the constituents from all directions, causing them to move randomly or explore the space. Such a space exploration is called diffusion. Diffusion enables the colloids to collide onto each other and thus interact.*

*The inter-colloids interaction determines how colloids self-assemble. For example, when the colloids are coated with complementary single-stranded DNA, these DNA strands hybridize into DNA bonds that link pairs of colloids together. In a macroscopic scale, these DNA-coated colloids have been reported to self-assemble into crystals whose structures were programmable. Scientists have run the simulations showing that by carefully designing the interactions between each pairs of colloids, one can direct the colloids to self-assemble into a mesoscopic Empire State Building. You can imagine such an ability will enable us to direct the colloids to self-assemble into needed components for micro-robots that may help us to fight against diseases such as performing treatments for clearing the clogs in the blood artery."*

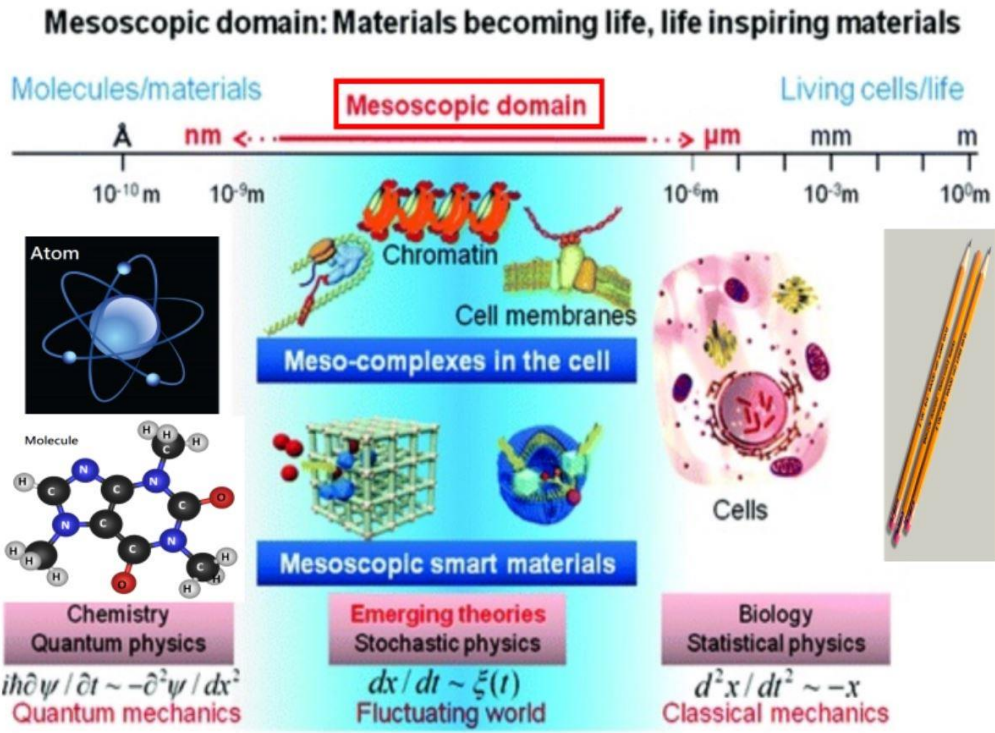

**Figure 3.1.3**: Illustration of the concept of mesoscopic scale (extracted from 00:50 of the video)

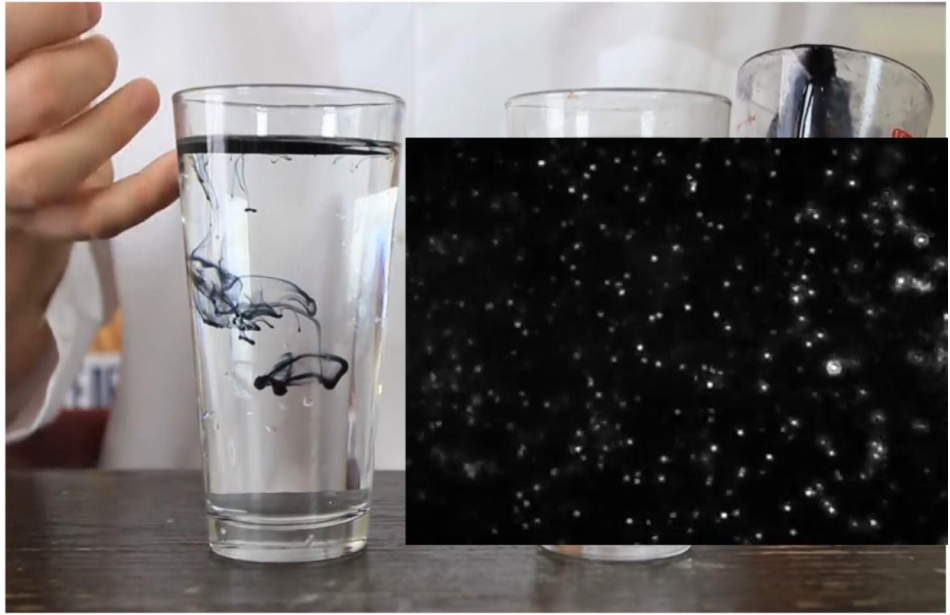

**Figure 3.1.4**: Illustration of the concept of diffusion (extracted from 01:09 of the video)

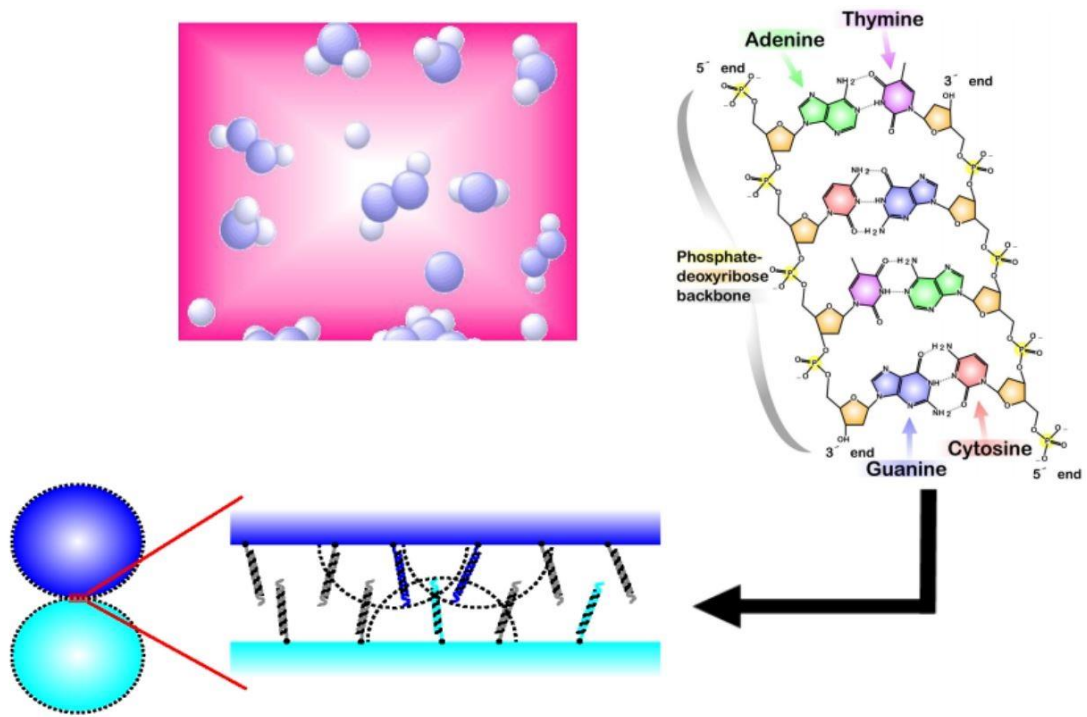

**Figure 3.1.5**: Colloids coated with complimentary single-stranded DNA and DNA strands would hybridize into DNA bonds that link pairs of colloids together (extracted from 01:44 of the video)

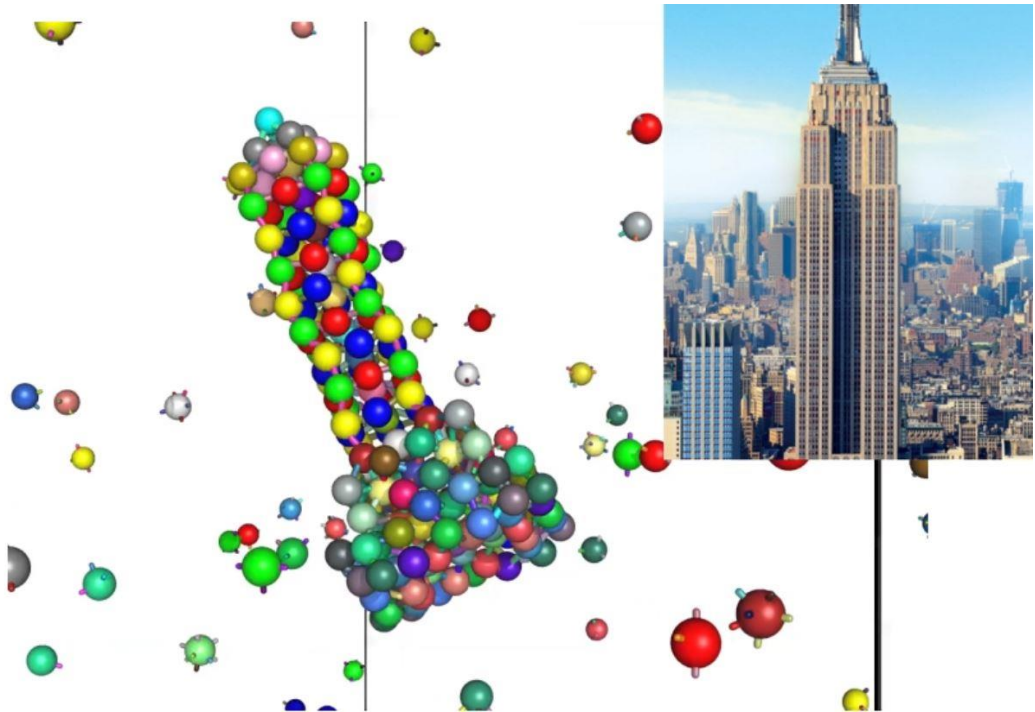

**Figure 3.1.6**: Simulation shows that colloids can be programmed to self-assemble into a mesoscopic empire state building (extracted from 02:11 of the video)

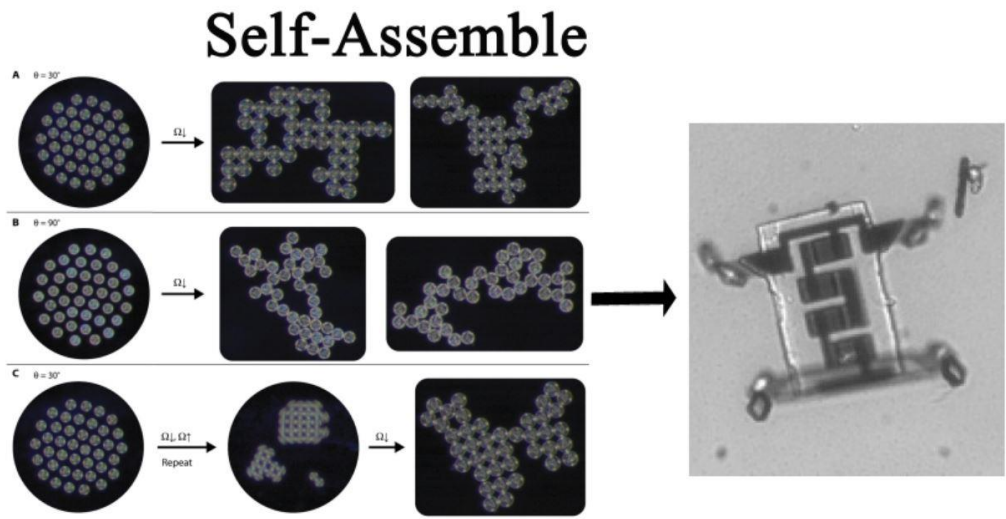

**Figure 3.1.7**: The programmable self-assembly of colloids can be applied to create micro-robots for medical purposes (extracted from 02:22 of the video)

In the third paragraph, we start introducing the concepts of active matter, which is a subfield of soft matter. First, we emphasize that active matter is different from passive matter by its capability of converting local chemical fuels to kinetic energy. Examples include bacteria consume sugar and oxygen to swim and colloidal surfers consume hydrogen peroxide to self-propel (***Figure 3.1.8***). Next, we start introducing the concept of active fluid, which comprises of active matter that stirs the surrounding fluid when self-propelling in a solution, as well as its most important physical property: self-organization. The example we use is microtubule-based active fluid which consists of pairs of microtubules bridged by kinesin motor clusters to slide apart (***Figure 3.1.9***). Here the sliding dynamics is accumulated from hundreds of microtubules to form a microtubule bundles that self-extend, stirs the surrounding fluid, and generate flows. Originally these flows form random vortices, canceling each other out in a long term. However, recent research indicates that by confining active fluid in circular containers such as cylinder or toroid regulates these flows to self-organize into a river-like coherent flow that circulates around the container (***Figure 3.1.10***). This phenomenon is critical on the applications of active fluid because not only it allows active fluid to transport materials in a distance beyond the limit of passive fluid, but it also demonstrates the potential of using the active fluid to perform mechanical work in mesoscopic system (***Figure 3.1.11***). To further impress our audiences on the importance of active fluid self-organization, we carry out one of the ultimate goals for active matter researchers, that is to create artificial cells because both the cytoplasm of eukaryotic cells and active fluid are powered by kinesin motor clusters which may share similar dynamics and control mechanisms (***Figure 3.1.12***). Therefore, the following contents are included in the third paragraph of the video:

*"One of the popular topics in soft matter is active matter. Active matter is differentiated from conventional passive matter due to its capability of converting local fuels to kinetic energy. For example, bacteria consume sugar and oxygen to swim; colloidal surfers consume hydrogen peroxide to self-propel. These active matter constituents interact to collectively form a macroscopic material flow. For example,*

*microtubule-based active fluid comprises of microtubules. These microtubules are bridged by kinesin motor clusters to slide apart. These sliding dynamics accumulate from hundreds of microtubules to form a microtubule bundles that self-extend. The bundle extension stirs the surrounding fluid, generating flows. These flows formed random vortices, canceling each other out in a long term. However, scientists have reported that confining this active fluid in a circular confinement such as cylinder or toroid regulates these flows into a river-like coherent flow, capable of transporting materials in a distance beyond the limit of passive fluid. This result demonstrates a potential of using the active fluid to perform mechanical work in mesoscopic system which serves as an alternative to power generation. Furthermore, by understanding the physical behaviors of active fluid, we can get a glimpse of how eukaryotic cells transport molecules through the fluidic cytoplasm powered by kinesin, which pushes us closer to the realization of artificial cells!"*

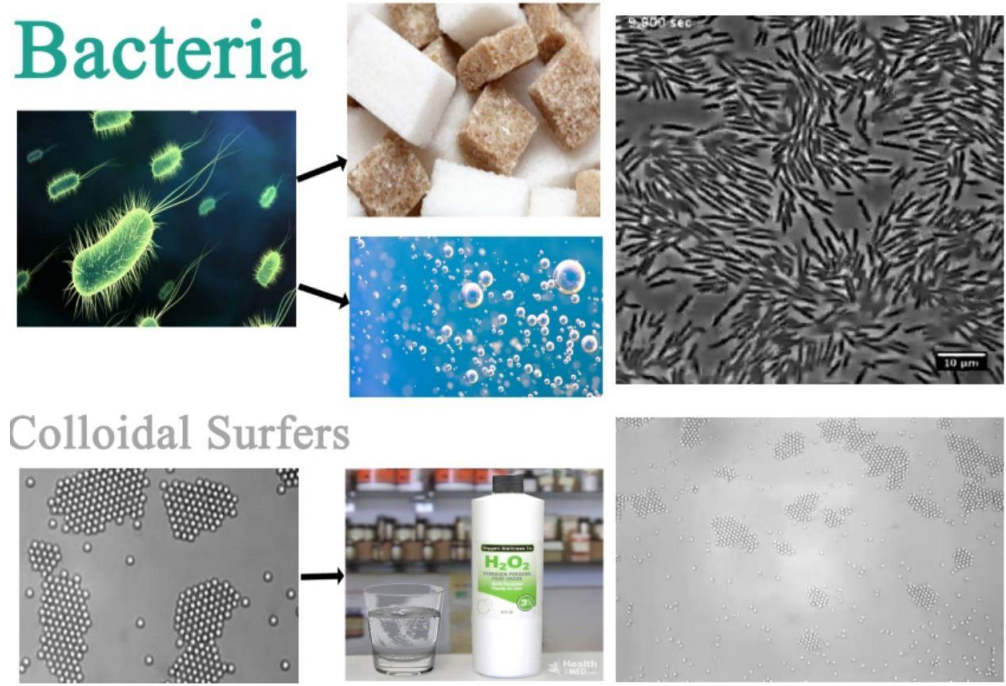

**Figure 3.1.8**: Active matter can convert local fuels to kinetic energy, such as bacteria consume sugar to swim; colloidal surfers consume hydrogen peroxide to self-propel (extracted from 03:00 of the video)

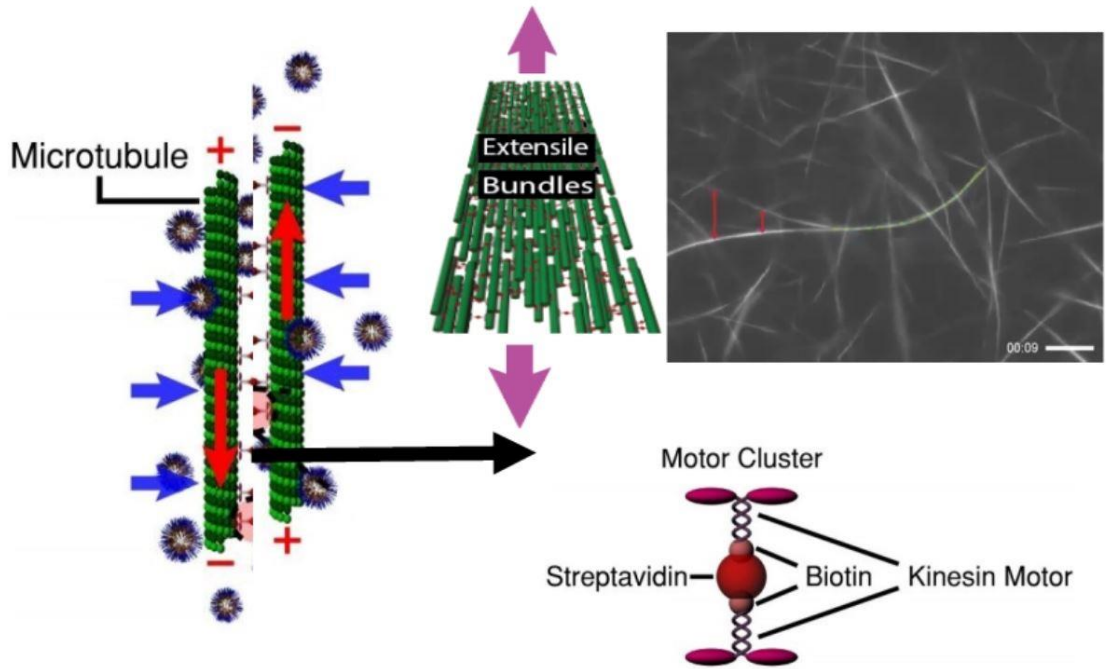

**Figure 3.1.9**: Mechanism of microtubule-based active fluid driven by kinesin motor clusters and self-extension of microtubule bundles (extracted from 03:25 of the video)

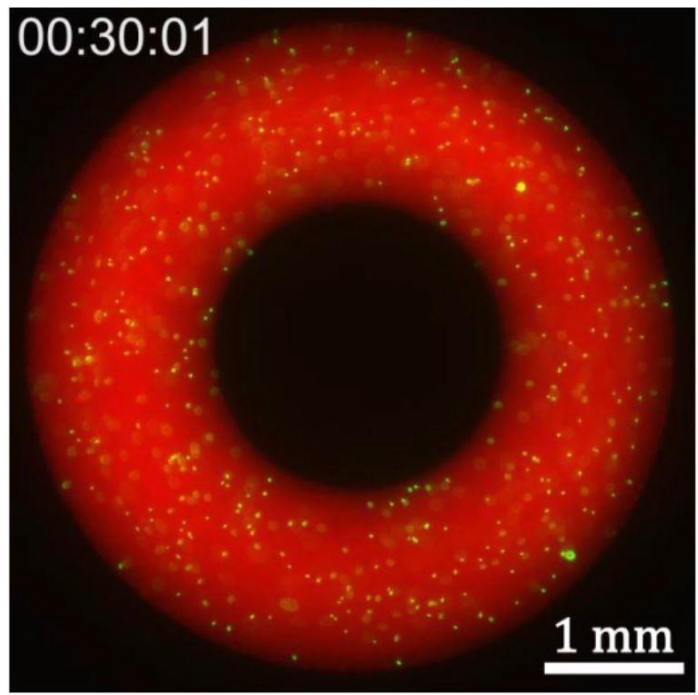

**Figure 3.1.10**: The active fluid self-organizes into a circulatory coherent flow when confined in circular containers (extracted from 03:46 of the video)

# Active Fluid Can Power a Microscopic Gear

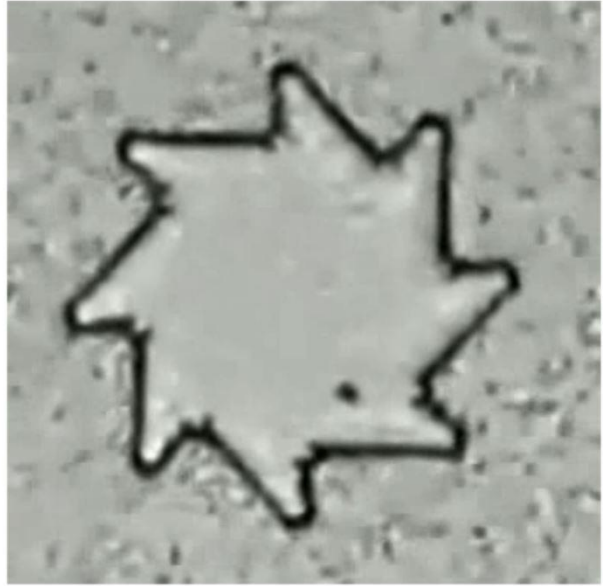

**Figure 3.1.11**: Active fluid can be used to perform mechanical work in mesoscopic systems (extracted from 04:15 of the video)

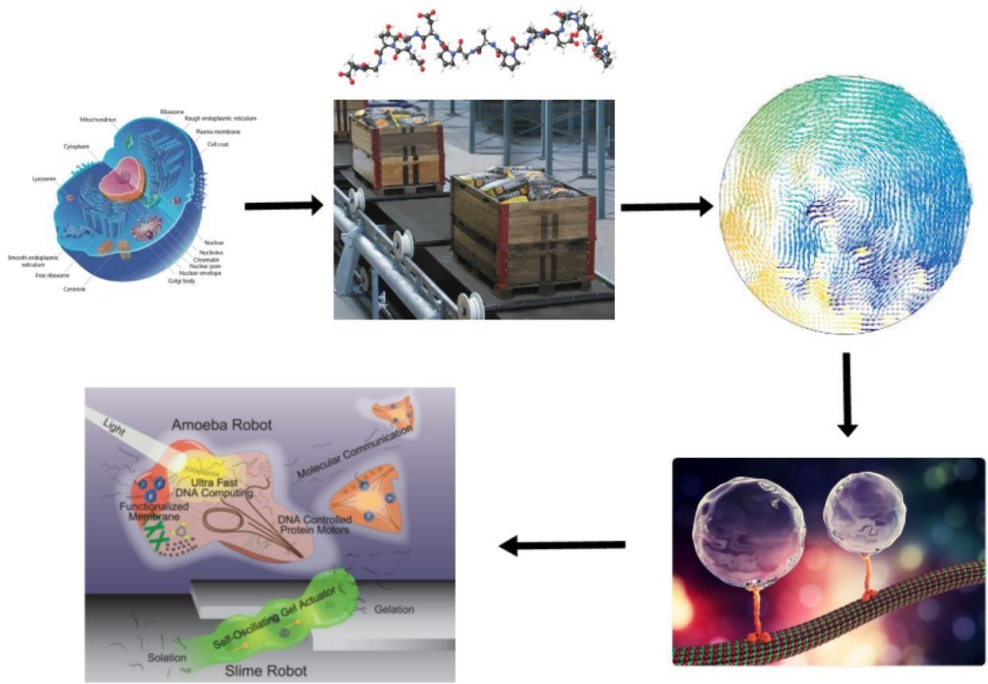

**Figure 3.1.12**: Research on active fluid power by kinesin motor clusters may help us to understand how eukaryotic cells transport molecules through the fluidic cytoplasm and thus push us closer to the realization of artificial cells (extracted from 04:30 of the video)

After introducing both the concepts of soft matter and active matter, we start introducing their potential real-life applications and well-paid career opportunities in the fourth paragraph. This is important because it is hard to attract people to study soft matter physics without economical attractions. First, we introduce the soft matter career opportunities in oil industries because some of the renowned oil corporations such as ExxonMobil and Solvay Rhodia are actively hiring soft matter scientists to improve the oil refinery process, such as extracting oil from water-oil mixture or understanding how oil goes through porous materials (***Figure 3.1.13***). According to the job searching page on Glassdoor, these jobs typically offer an annual salary ranging from $100K to $130K which is way higher than other job opportunities in other industries. Next, we want our audiences to understand that soft matter career opportunities are not limited to oil industries, rather, they are everywhere in the production of our daily products. The examples we introduce include cosmetic, cleaning, and designing microfluidic medical devices that can select the most active sperms to resolve the aggravating infertility rate nationwide (***Figure 3.1.14***). Lastly, we impress our audiences again about the potential economic growth on soft matter industries by claiming that one who has successfully developed and commercialized that sperm selection device may become a multibillionaire. Therefore, the following contents are included in the fourth paragraph of the video:

*"While soft matter and active matter have demonstrated their rich academic value, they also lead to miscellaneous industrial applications and thus career opportunities. There are corporations who are hiring people with experience in soft matter. For example, the oil companies such as ExxonMobil and Solvay Rhodia demand soft matter scientists to improve the oil refinery process, such as extracting oil from water-oil mixture or understanding how oil goes through porous materials. According to Glassdoor, these jobs typically offer an annual salary ranging from $100K to $130K. The background in soft matter also allows you to work in cosmetic companies such as L'Oréal or Unilever and cleaning companies such as Paul Mitchell or Garnier because cosmetics and shampoos are both part of soft matter. Developing*

*these products requires a background in soft matter research. Soft matter also plays an important role in medical health. For example, according to Key Statistics from the National Survey of Family Growth (NSFG), from 2015 to 2017, there are 13.1 percent of women in the United States aging from 15 to 44 years old have the problem of infertility. Therefore, developing a microfluidic device that can efficiently select active "sperms" has become a fast-growing business. Designing this device involves the knowledge of controlling the fluid at "mesoscopic" scale, thus within the scope of soft matter. If you are a soft matter scientist and successfully develop and commercialize such a device, you will become a multi-billionaire."*

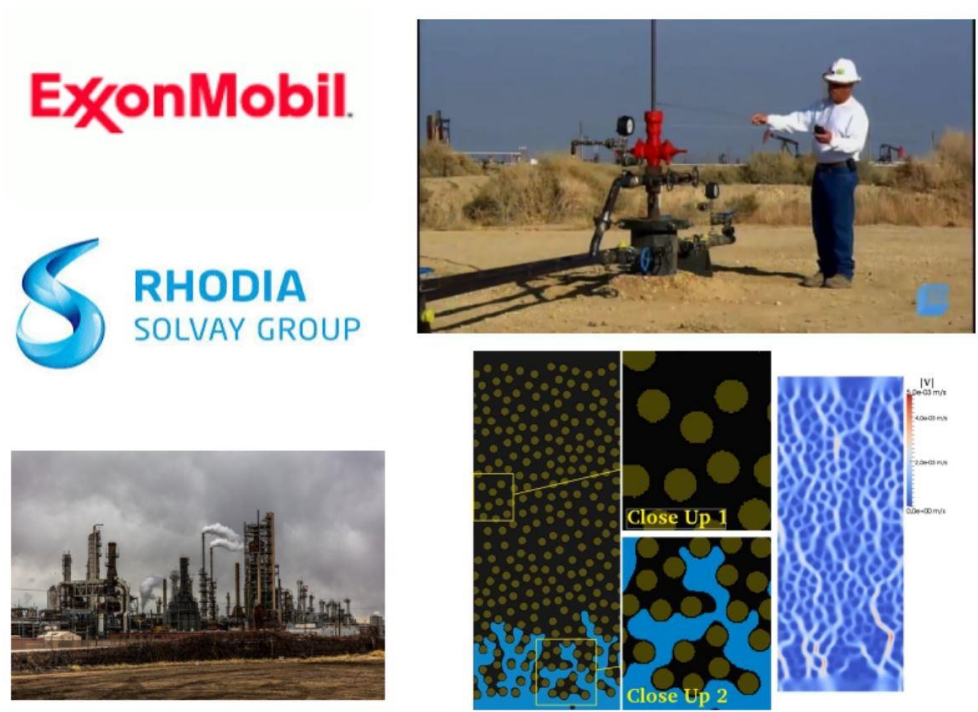

**Figure 3.1.13**: Oil companies such as ExxonMobil and Solvay Rhodia are actively hiring soft matter scientists to improve the oil refinery process, such as extracting oil from water-oil mixture or understanding how oil passes through porous materials (extracted from 05:07 of the video)

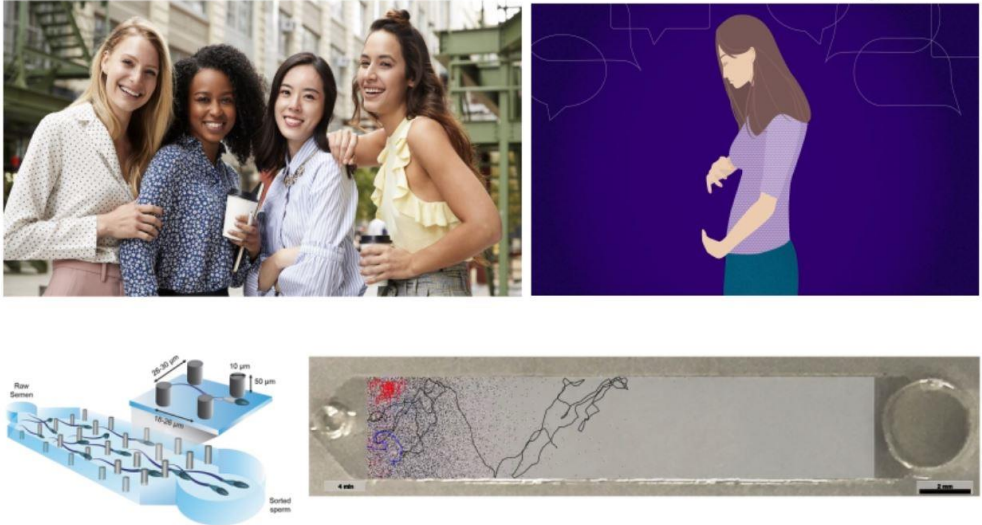

**Figure 3.1.14**: Research on soft matter and active matter may be applied to develop sperm selection device that helps reduce the infertility rate nationwide (extracted from 06:24 of the video)

To conclude the video, we first emphasize again that soft matter is everywhere in our life and thus research works on soft matter, including synthesizing new materials at mesoscopic scale and the sperm selection device that reduces infertility rate, are critical on creating a better life in the future (***Figure 3.1.15***). Next, we mention again the well-paid career opportunities and the rapid growth of soft matter industries and thus encourage ambitious students to learn more about soft matter physics and be involved in the related industries after they have graduated from college. Therefore, the following contents are included in the last paragraph of the video:

*"In summary, soft matter are ubiquitous in our life. From a rain drop in the sky to the food in your lunch box are all parts of soft matter. The soft matter research has not only shown the feasibility of synthesizing new materials at mesoscopic scale but also demonstrated its potential to help many couples*

*to complete their families. Soft matter also leads to well-paid career opportunities in industries ranging from oil to cosmetic and house cleaning. New industries and businesses are booming to produce products using technologies in soft matter. This is the field full of opportunities and challenges, allowing you to use your talent to develop new technologies as well as advance our understanding of the science in our everyday life."*

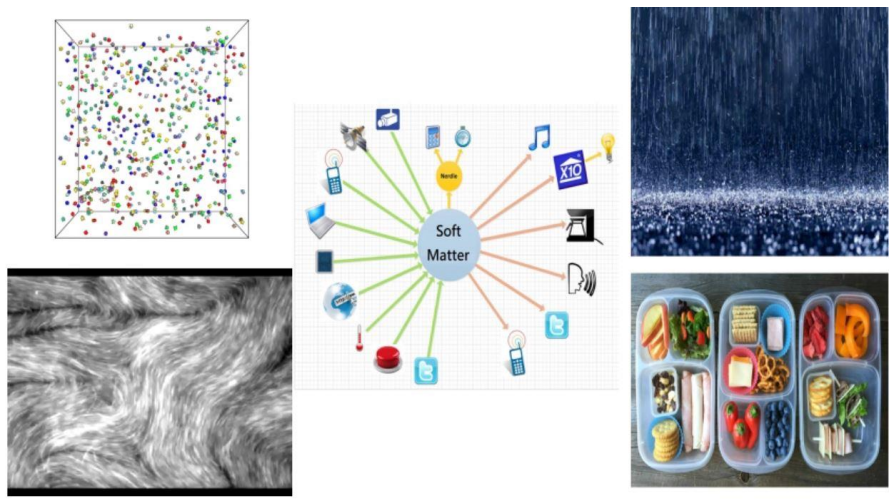

**Figure 3.1.15**: To illustrate the concept that soft matter is ubiquitous in our daily living in the conclusion of the video (extracted from 06:46 of the video)

Although this video presents complete concepts of soft matter and active matter, one of its disadvantages is that the 7.5 minutes total length may be too long for some people to be concentrate on. Hence, we also make a short version of this video which excludes all the active matter contents (paragraph 3 and part of paragraph 4 that involves the sperm selection device) as well as the conclusion paragraph. This makes the total length of the short version video to be roughly 3.5 minutes, which should increase the possibility that audiences finish watching the video. Below are the links for finding both videos on YouTube:

Full version: https://youtu.be/ga2e8XEkuR8

Short version: https://youtu.be/kbdQ_yHvvUk

## 3.2: Design of Online Survey

In this study, we used the Google Form to conduct online survey on assessing the student or public awareness, understanding as well as personal interests in soft matter physics and related research or career opportunities before and after watching our short video described in the previous section. Below I will briefly illustrate the contents of the survey as well as their functions on the educational assessments and the promotion of soft matter educations or careers.

The entire survey is split into five pages and is named "Survey on Student Understanding to Soft Matter." In the beginning of page 1 (***Figure 3.2.1***), we first make a short introduction on the purposes and the estimated time it may take to finish our survey in order to encourage people to answer the survey questions. Below is the content of short introduction:

*"Please spend ~5 minutes answering the following questions HONESTLY. This survey is meant to assess the public understanding of soft matter along with associated industrial applications and career opportunities. Your response is confidential and will be used as an important assessment for the need to offer related educational resources at WPI."*

**Figure 3.2.1**: The first page asking about the respondents' general information, including current school year, major, and highest degree they plan to pursue

After introducing the survey purpose, we ask three questions on the first page of survey, including school year, major, and the highest degree the respondents plan to pursue. This is meant to understand the respondents' ages, academic plans and interests.

On the second page of the survey (*Figure 3.2.2*), we ask our respondents to indicate the levels of interests from 1 to 10 (1 means no idea or not interested at all; whereas 10 means strong interest) to multiple academic disciplines, including physics, mathematics, chemistry, biology, biochemistry, biophysics, and material science. This aims to evaluate respondents' current academic interest as well as to analyze the relation between academic interest and the awareness of soft matter related knowledge. In order to assist our respondents to answer the questions on this page, we made the following illustration at the beginning of the page:

*"Please indicate your level of interest and understanding to the following disciplines (1: No idea at all; 10: I like it strongly)"*

On the third page of the survey (**Figure 3.2.3**), we start evaluating respondents' awareness and interest in soft matter related knowledge, research, applications, and career opportunities before watching our video by asking respondents to indicate their levels of agreement to several statements (1 means not at all whereas 10 means strongly agree). In order to assist the respondents to answer the questions on this page, we made the following illustration at the beginning of the page:

*"The following questions are for collecting the participant's background in science. Please answer the following questions before watching the video (1: Strongly Disagree; 10: Strongly Agree)"*

The questions include:

*"I am familiar with biophysics"*

*"I am familiar with soft matter"*

*"I am aware of real-life applications of soft matter"*

*"I am aware of the soft matter research in WPI"*

*"I have performed the soft-matter related research"*

*"I am aware of soft-matter related career opportunities"*

At the end, we ask the respondents to name a few examples of career opportunities in soft matter to evaluate whether they understand the job offered in the field of soft matter as well as whether they answer the last question honestly.

**Figure 3.2.2**: The second page of the survey asking about the respondents' academic interest

**Figure 3.2.3**: The third page of the survey asking about the respondents' interest and understanding to soft matter before watching the video

On the fourth page of the survey (***Figure 3.2.4***), we ask our respondents to watch the videos we make as illustrated in Section 3.1 of this report. Here, we offer both the full and short versions of the video and let the respondents to decide which video to watch.

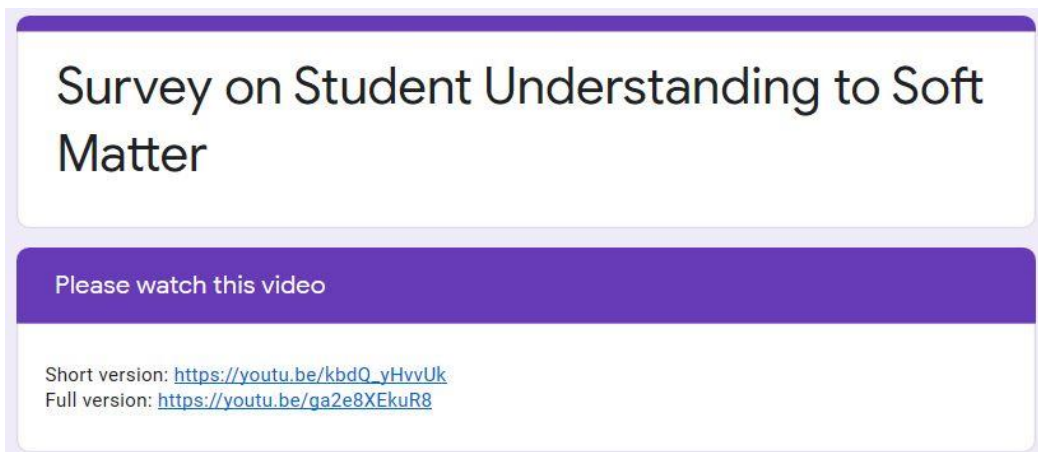

**Figure 3.2.4**: The fourth page of the survey asking the respondents to watch the video about soft matter

On the last page of the survey, we first generally ask our respondents to indicate their levels of awareness and interest in soft matter research and applications after watching our video (1 means not interested at all whereas 10 means strong interest and awareness; ***Figure 3.2.5* A**). This will help us to evaluate whether the video has effectively increased people's awareness on soft matter. We also ask the respondents to name a few professors at WPI who conduct research on soft matter, which indicates whether the respondents start to pay attention to soft matter research at least at their own college. The questions include:

*"My awareness of soft matter"*

*"My awareness about the real-life applications of soft matter"*

*"My awareness of the soft matter research in WPI"*

*"Please name the WPI professors you know who are working on soft matter"*

*"My interest of performing the soft-matter related research"*

*"My interest of pursuing soft-matter related career"*

Next, to further evaluate the effectiveness of video to stimulate students' interest in different aspects of soft matter research and career opportunities, we designed a question that allows the respondents to rank their impressions on five important concepts mentioned in the videos (***Figure 3.2.5 B***). These include:

*"The promising capability of self-assembling into a designed structure such as mesoscopic Empire State Building"*

*"Fruitful academic values such as in DNA hybridization process, mechanism of entropic forces, non-equilibrium physics, and mixing dynamics of self-pumping fluids"*

*"Soft matter is ubiquitous in our everyday life, ranging from food, fossil fuels to cosmetics and TV"*

*"Potential medical applications such as microrobots"*

*"Rich career opportunities"*

*"Others"*

Lastly, we ask our respondents that how much and how should the government and colleges support the education and research on soft matter, including the level of difficulties and the departments to provide such courses (***Figure 3.2.5 C***). This helps us to further evaluate whether the respondents are persuaded that the solid college education is pivotal on the booming of soft matter industries and research and whether soft matter education can lead to a good career path in the industries. These are important factors to encourage more students to dedicate themselves to the soft matter related industries. The questions and options include:

*"I think the government should increase the funding in soft matter research to stimulate the growth of related industries"*

*"I think WPI should open a course that teaches soft matter"*

*"What level should the course be taught?"*

1. *Introductory (1000 Level)*
2. *Intermediate (2000 Level)*
3. *Advanced (3000 or 4000 Level)*
4. *Graduate (5000 or above)*
5. *Others*

*"What department should provide such a course? Please select all that apply*

*(Physics, Mechanical Engineering, Biomedical Engineering, Chemistry and Biochemistry, Chemical Engineering, Electrical Engineering, Computer Science, Mathematics, Humanity, Others)"*

**Figure 3.2.5**: The fifth page of the survey asking about (**A**) Respondents' interest and understanding to soft matter after watching the video (**B**) Respondents' impression on five important concepts in the video (**C**) Respondents' opinion on whether government should increase funding on soft matter research and whether college should open soft matter courses as well as how to distribute these resources in different school years and departments

It is noted that in order for us to receive more replies to the survey and increase the validity of the statistics and analysis, we not only send this survey to students at Worcester Polytechnic Institute (WPI), but we also send it to several other universities, including National Taiwan University (Taipei, Taiwan), National Cheng Kung University (Tainan, Taiwan), as well as the general public. We expect to receive at least 50 replies in order to let the statistics on survey results to be meaningful, which will be described in detail in the next section.

## 3.3: Statistics and Analysis on Survey Results

In this paper, several statistics will be done by MATLAB on the survey replies in order to evaluate the effectiveness of promoting soft matter knowledge and education via the homemade video as described in Section 3.1. Before conducting the statistics, we would like to define several commonly used statistical terms in this study. First, the average ($\bar{x}$) is defined as

$$\bar{x} = \frac{1}{n} \sum x_n \tag{1}$$

where n is the number of samples.

The standard deviation ($\sigma$) is defined as

$$\sigma = \sqrt{\frac{\sum (x_n - \bar{x})^2}{n-1}} \tag{2}$$

Next, the student impression factor ($IF$) on different concepts mentioned in the video (page 5 of the survey) is calculated based on the order of respondents' impression defined as

$$IF = \frac{\sum (6 - S)}{n} \tag{3}$$

where S is the order of impression from each respondent on each concept. S = 1 when the concept is the respondent's first impression and so on; S = 6 when the concept is the respondent's last impression.

Finally, the improvement factor ($\varepsilon$) on the average level of student knowledge and interest on soft matter before ($\overline{x_b}$) and after ($\overline{x_a}$) watching the video is defined as

$$\varepsilon = \overline{x_a} - \overline{x_b} \tag{4}$$

The first question to be answered by the survey is that whether the video has effectively stimulated students' interest and understanding to biophysics and soft matter as well as pursuing related research and careers. To answer this question, we compare the average of people's levels of interest or understanding to soft matter as well as the related research and career opportunities before (page 3 of survey) and after (page 5 of the survey) watching the video. In addition, the standard deviation will be calculated to evaluate whether the knowledge and interest gaps on soft matter between each respondent have shrank after watching the video. To further understand which concepts in the video have impressed the audiences the most, the student impression factor (IF) on different concepts will be calculated and compared. This result will be useful on the design of soft matter courses in order to satisfy students' interest and future needs.

The second question to be answered is that whether students with different majors, school ages, highest degrees planned to pursue have different levels of understanding and personal interest in soft matter related research and career opportunities before and after watching the video. To answer this question, we first sort all the majors of respondents into 4 main categories: Basic Science (mathematics and physics), Physical Engineering (mechanical engineering, electrical engineering, aerospace engineering, computer science, and hydraulic and ocean engineering), Chemical and Bio-engineering (chemical engineering, chemistry and biochemistry, biomedical engineering, medical school, and dentistry), and Social Study (music, geology, Chinese, business, management, and politics). This way we can ensure that each category has more than 10 replies from the survey in order to be considered as a meaningful statistic. Next, we will compare their average and standard deviation on students' levels of knowledge and interest in soft matter before and after watching the video. The same process will be done for addressing the effects on school ages and highest degree planned to pursue.

The third question to be answered is that the relation between students' levels of interest on different subjects (page 2 of the survey) to their knowledge and interest in soft matter before and after watching the video. To answer this question, at every 2 levels of interest of different subjects (which are 2, 4, 6, 8, and 10), we calculate the improvement factor on the levels of student knowledge and interest in soft matter before and after watching the video. It is noted that the improvement factor calculated here is the average of the answers to 5 questions related to soft matter knowledge and interest before and after watching the video, which are on the third and fifth page of the survey respectively.

The last question to be addressed is that whether WPI should open a course focusing on soft matter and biophysics as well as how to distribute the educational resources in different school ages and departments. To answer this question, we first calculate the average levels of demand of government resources on soft matter research and the offer of college soft matter courses (page 5 of the survey) after watching the video. In addition, the standard deviation will be calculated to address whether all the respondents have consensus to these survey questions. Next, we will evaluate what is the most suitable school years as well as departments to provide soft matter and biophysics courses by counting the votes on the related survey questions located on the 5[th] page of the survey. It is noted that the above statistics will be conducted for the overall respondents as well as students with different majors and school ages for further comparison.

# Chapter 4: Results and Discussion

Until March 06, 2020, 65 replies have been received from the survey (37 replies from Worcester Polytechnic Institute; 4 from National Taiwan University; 4 from National Cheng Kung University; 20 from the general public), which have satisfied the minimum requirement on the amount of replies, which is 50 as described in Section 3.2. Therefore, the four different statistics as mentioned in Section 3.3 have been completed in order to address the effectiveness of promoting soft matter knowledge and education via the homemade video.

## 4.1: Overall Effectiveness of Video on Boosting Soft Matter Interest

The first statistic aims to answer whether the video has effectively stimulated students' understanding and interest in pursuing soft matter related research and career opportunities as well as to determine which concepts in the video have impressed the audiences the most. Thus, **Figure 4.1.1** shows the average of people's levels of interest and understanding to soft matter before and after watching the video for 5 related questions as shown on the 3$^{rd}$ and 5$^{th}$ page of the survey, which are "My awareness of soft matter", "My awareness about the real-life applications of soft matter", "My awareness of the soft matter research", "My interest of performing the soft-matter related research", and "My interest of pursuing soft-matter related career". The figure indicates that after watching the video, the respondents' levels of interests in soft matter related research, applications, and careers have increased by 1.35 to 3.51 points (with average increment of 2.78 points out of 10; 27.8 %), which shows that the video has stimulated people's awareness on soft matter. We can also see that the standard deviations on people's interest or awareness to soft matter have decreased after watching the video. This indicates that the soft matter knowledge gaps between each respondent have been shrank by watching the video. In addition, the figure shows that people's increment of interest in performing soft matter research or related careers after watching the video is comparatively lower than those of acquiring information on soft matter related applications and research. This can be reasonably explained because people are willing to perform research on certain fields or be committed to certain careers only if they have strong personal interest or background knowledge, which is nearly impossible to be provided in a 7-minute video.

To further determine which concepts in the video have impressed the audiences the most, ***Figure 4.1.2*** indicates the student impression factor (IF) for 5 major soft matter concepts mentioned in the video, which are **Concept 1**: The promising capability of self-assembling into a designed structure such as mesoscopic Empire State Building; **Concept 2:** Fruitful academic values such as in DNA hybridization process, mechanism of entropic forces, non-equilibrium physics, and mixing dynamics of self-pumping fluids; **Concept 3**: Soft matter is ubiquitous in our everyday life, ranging from food, fossil fuels to cosmetics and TV; **Concept 4**: Potential medical applications such as microrobots; **Concept 5**: Rich career opportunities; and **Others**. From the figure we can see that the top three concepts that impressed all the audiences the most are Concepts 1, 2, and 5. These will be the three major points to be addressed in the future soft matter courses as they have covered both the explanation of soft matter physical behaviors and phenomena as well as its real-life applications and well-paid career opportunities. These will be considerably helpful on attracting more students to get involved in soft matter related knowledge, research, and careers.

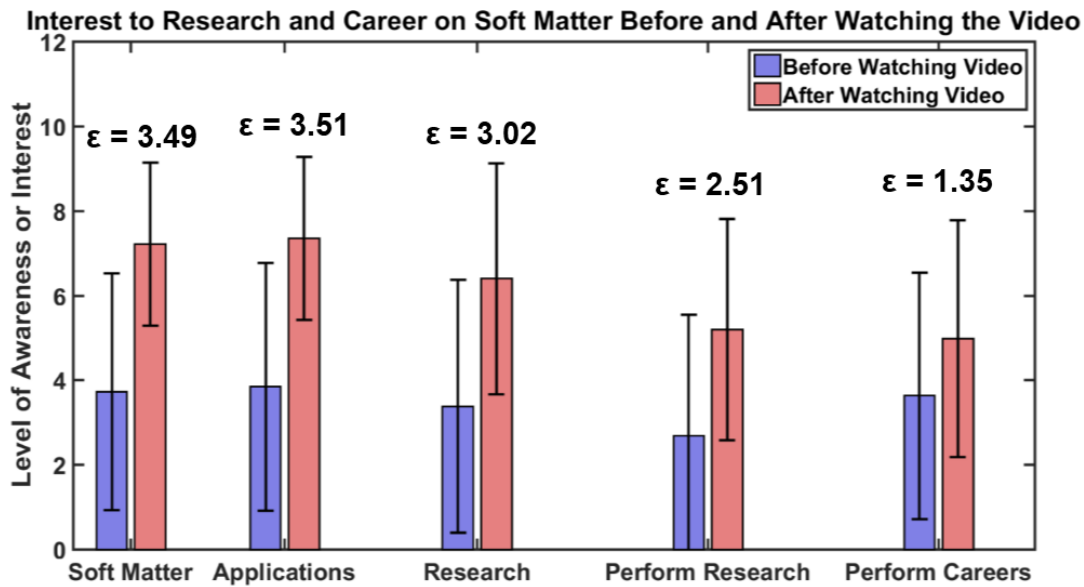

**Figure 4.1.1**: Level of awareness and interest in **general soft matter**, **soft matter real-life applications**, **soft matter research**, **performing soft matter research**, **performing soft matter careers** as well as the difference before and after watching the video

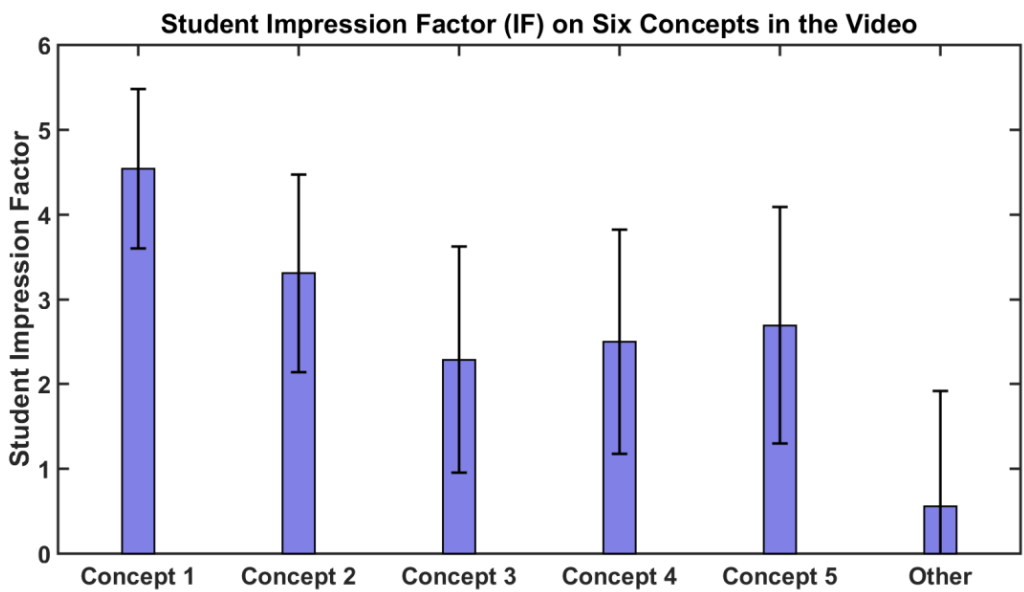

**Figure 4.1.2**: Student impression factor (IF) on 5 different concepts mentioned in the video, which are ***Concept 1***: The promising capability of self-assembling into a designed structure such as mesoscopic Empire State Building; ***Concept 2***: Fruitful academic values such as in DNA hybridization process, mechanism of entropic forces, non-equilibrium physics, and mixing dynamics of self-pumping fluids; ***Concept 3***: Soft matter is ubiquitous in our everyday life, ranging from food, fossil fuels to cosmetics and TV; ***Concept 4***: Potential medical applications such as microrobots; ***Concept 5***: Rich career opportunities; and ***Others***.

## 4.2: Effects of Majors, School Years, and Degrees on Soft Matter Interest

The second statistic aims to clarify whether students' levels of interest and understanding to soft matter before and after watching the video differs from their majors, school ages, and the highest degrees they plan to pursue. Therefore, *Figure 4.2.1* shows the average levels of interest and understanding to soft matter before and after watching the video for four different majors, which are (A) Basic Science (mathematics and physics), (B) Physical Engineering (mechanical engineering, electrical engineering, aerospace engineering, computer science, and hydraulic and ocean engineering), (C) Chemical and Bio-engineering (chemical engineering, chemistry and biochemistry, biomedical engineering, medical school, and dentistry), and (D) Social Study (music, geology, Chinese, business, management, and politics). It is obvious to see that regardless of student majors, after watching the video, the students' interest and understanding to soft matter have increased whereas the standard deviation has decreased and thus the soft matter knowledge gaps between each student. Among all majors of studies, students who pursue chemical and bio-engineering related majors have the highest degree of interest and understanding to soft matter after watching the video whereas students who perform social study related majors have the lowest. This is because the physical properties of soft matter involve considerable amount of knowledge related to its molecular structure as well as the programmable self-assembly of particles to crystals are directly related to chemical and biomedical engineering. It is also noted that students who pursue physical engineering related majors have the largest improvement on the levels of interest and understanding to soft matter after watching the video because their knowledge in soft matter is comparatively insufficient before watching the video. This is because courses related to biophysics, biochemistry, and soft matter are seldomly required or even offered within the current engineering program curriculum, which means by offering soft matter courses to those students, one may inspire their interest to express the field of soft matter by either conducting scientific research or pursuing related careers.

Next, *Figure 4.2.2* shows the average levels of interest and understanding to soft matter before and after watching the video for four different school years, which are (A) First and Second Year

Students, (B) Third and Fourth Year Student, (C) Graduate Students, and (D) Faculties. It is obvious to see that regardless of student school years (faculties excluded), after watching the video, the students' interest and understanding to soft matter have increased whereas the standard deviation has decreased and thus the soft matter knowledge gaps between each student. Among all student school years, if faculties are excluded, third- and fourth-year undergraduate students have the highest levels as well as the greatest improvement on the interest and understanding to soft matter after watching the videos compares to other school years. This implies that offering introductory soft matter courses to third- and fourth-year undergraduate students would be the most appropriate choice as it is the easiest to boost their interests on soft matter related concepts and phenomena, such as programmable self-assembly driven by diffusion of depletant and the self-organization of microtubule-based active fluid, because they already have adequate background knowledge to understand these concepts, which may be too hard for first- and second-year students but too easy for graduate students. This can also be seen from the statistical results of graduate students and faculties as those people have comparatively low improvement on their interest and understanding to soft matter after watching the video because they may already have some basic understanding to soft matter physics before watching the video. The same situation happens to first- and second-year students as their background knowledge may be inadequate to absorb the information related to the mechanisms of soft matter.

Finally, ***Figure 4.2.3*** shows the average levels of interest and understanding to soft matter before and after watching the video for three different highest degrees students planned to pursue, which are (A) Bachelor, (B) Master, and (C) PhD. It is obvious to see that regardless of the highest degrees students planned to pursue, after watching the video, the students' interest and understanding to soft matter related information have increased to a similar level whereas the standard deviation has decreased and thus the soft matter knowledge gaps between each student. This indicates that the video has effectively stimulated students' interest in soft matter regardless of the highest degrees they plan to pursue, which is an inspiring result for the future soft matter courses providers because we have proved that the concepts and phenomena related to soft matter are not only inspiring but also understandable to mostly everyone. However, we can see that students who plan to pursue PhD degree have the higher levels of interest on performing soft matter research and careers after watching the video compared to those who only plan to pursue master or bachelor's degree. This can be explained since PhD degree is research oriented, and students who plan to receive this degree should have the determination on conducting research in depth in a certain field.

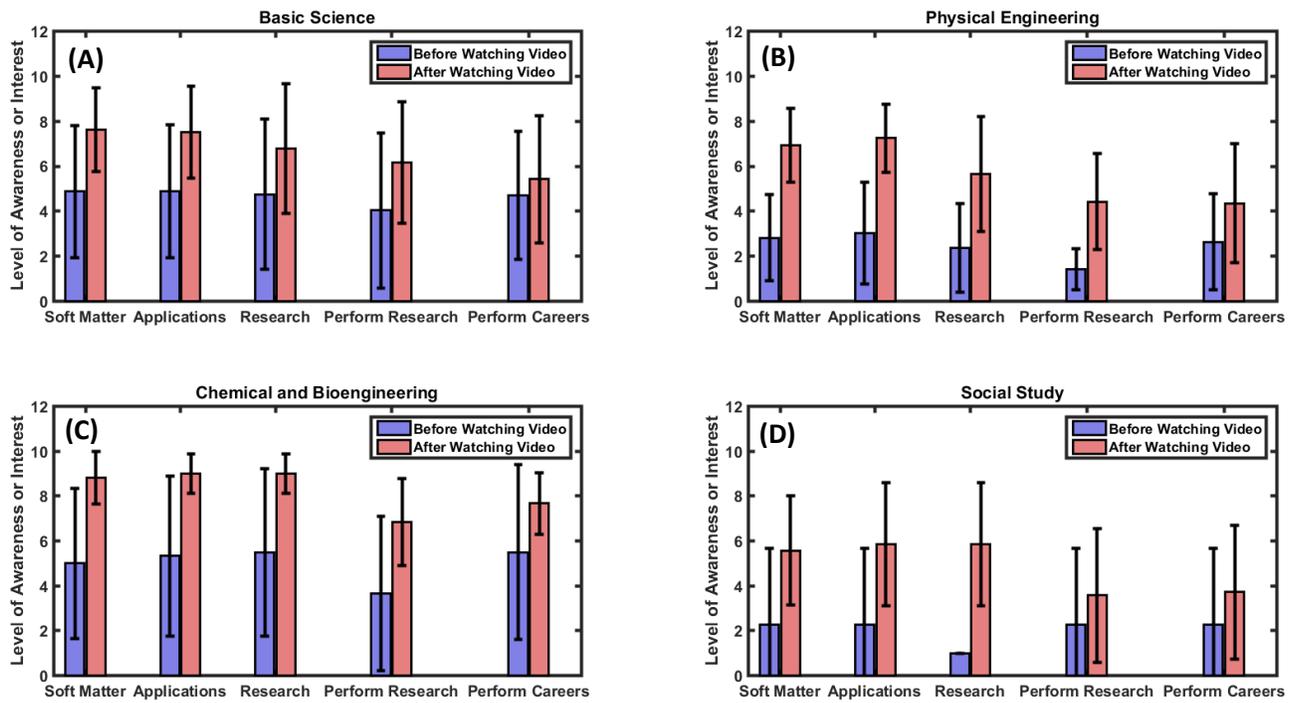

**Figure 4.2.1**: Levels of awareness and interest in soft matter related knowledge, research, and careers for students pursing four different majors before and after watching the video, which are **(A)** Basic Science (mathematics and physics), **(B)** Physical Engineering (mechanical engineering, electrical engineering, aerospace engineering, computer science, and hydraulic and ocean engineering), **(C)** Chemical and Bio-engineering (chemical engineering, chemistry and biochemistry, biomedical engineering, medical school, and dentistry), and **(D)** Social Study (music, geology, Chinese, business, management, and politics)

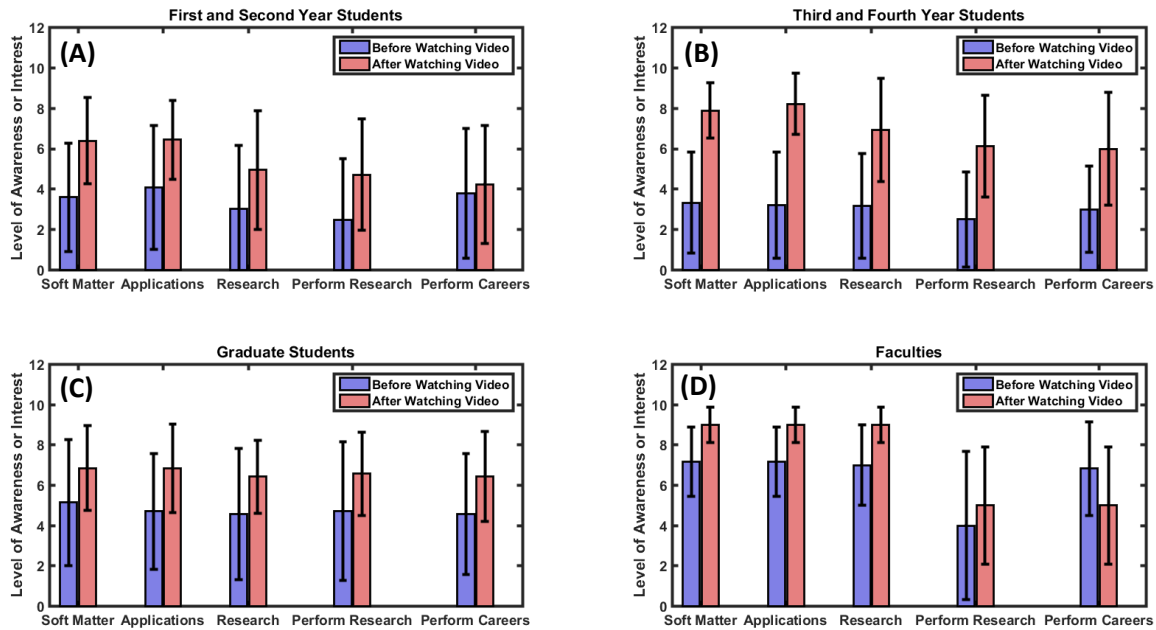

**Figure 4.2.2**: Levels of awareness and interest in soft matter related knowledge, research, and careers for students pursing four students with four different school ages before and after watching the video, which are **(A)** First and Second Year Students, **(B)** Third and Fourth Year Student, **(C)** Graduate Students, and **(D)** Faculties

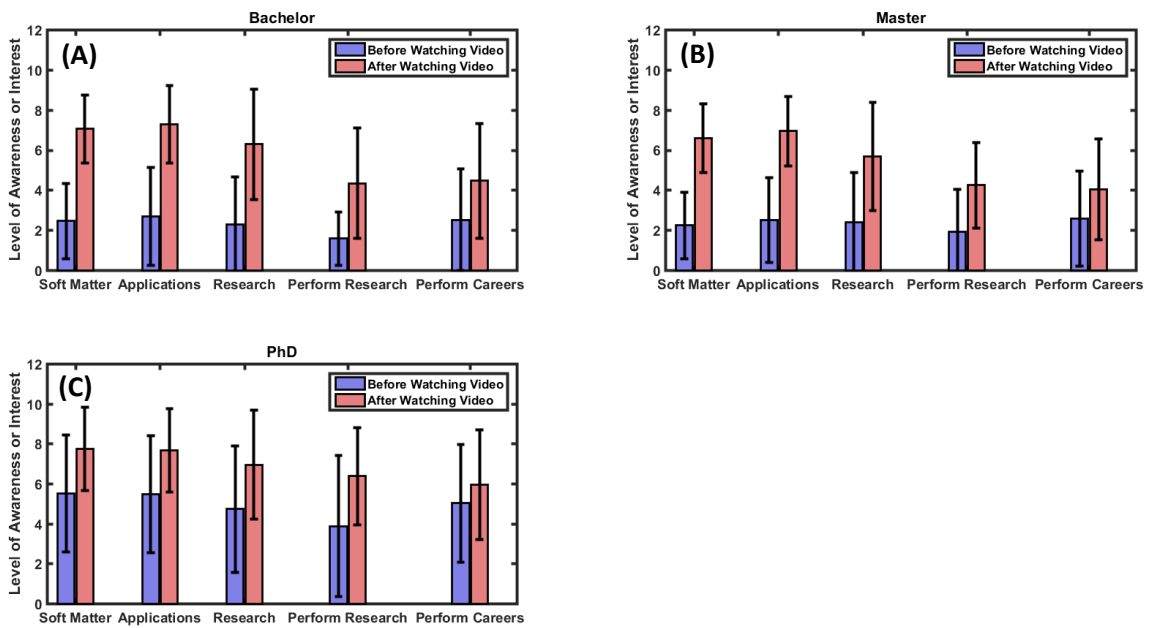

**Figure 4.2.3**: Levels of awareness and interest in soft matter related knowledge, research, and careers for three different highest degrees students planned to pursue, which are (A) Bachelor, (B) Master, and (C) PhD

### 4.3: Effects of Subjects Interest Levels on Soft Matter Awareness

The third statistic aims to clarify the relation between students' levels of interest on different subjects to the improvement of their interest and knowledge in soft matter after watching the video. Therefore, ***Figure 4.3.1*** shows the average improvement factor (ε) of 5 questions related to soft matter knowledge and interest after watching the video under different levels of interest of different subjects. It is noted that the improvement factor is calculated for every two subject levels of interest in order to have more samples included and provide a more stable statistical result. We can see that out of seven different subjects, the subject levels of interest have three types of effects on the average improvement factor of students' awareness to soft matter. The first type of trend includes the subject of **(A)** physics, biophysics, and material science which the average improvement factor reaches the maximum when the subject levels of interest range from 4 to 6 and decrease to nearly 0 when the subject levels of interest approach 10. This can be reasonably explained since these three subjects are directly related to soft matter and people who have higher levels of interest to these subjects should have learned the soft matter related knowledge before watching the video. Thus, students with smaller levels of interest to these three subjects tend to learn substantially about soft matter after watching the video, which leads to higher improvement factors.

The second type of trend includes the subject of **(B)** mathematics and biochemistry which the average improvement factors first decrease then increase as students' levels of interest to these two subjects increase (the critical point occurs when the subject levels of interest equal to 6). This may be explained by the fact that these two subjects are partially related to soft matter, such as the coating of DNA colloids, mixture of active fluid, and the programming of colloidal self-assembly. Therefore, students who have extremely low or extremely high interest in these subjects tend to be inspired more by the video because either they have completely no idea on soft matter before watching the video or soft matter related concepts mentioned in the video have inspired them to start up the related research in their own fields of study.

The third type of trend includes the subjects of **(C)** chemistry and biology which the average improvement factors are all positive and does not affected by the students' level of interest to these subjects. This may be due to the fact that soft matter is not directly related to these two fields so that the related concepts and phenomena such as self-assembly and self-organization remain new and inspiring to students who study in these fields.

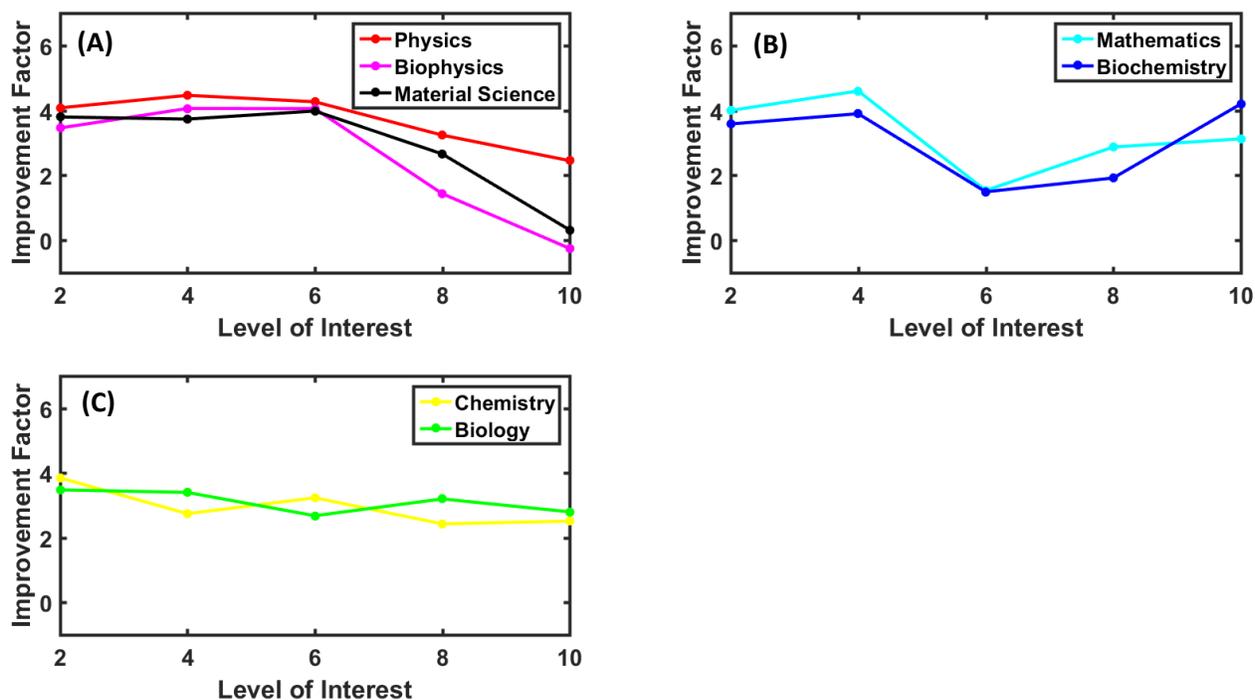

**Figure 4.3.1**: Average improvement factor ($\varepsilon$) on students' interest and understanding to soft matter under different levels of interest of different subjects, including **(A)** Physics, Biophysics, Material Science; **(B)** Mathematics, Biochemistry; **(C)** Chemistry, Biology

## 4.4: Offer of Soft Matter Courses and Educational Resource Distribution

Finally, the fourth statistics aims to clarify whether WPI should invest more financial resources on offering soft matter courses and supporting the related research as well as how to distribute these resources to different school ages and departments. Therefore, ***Figure 4.4.1*** shows the levels of agreement on more state funding to soft matter research and the offer of college soft matter courses after watching the video responded by students with different majors **(A)** and school years **(B)**. It is obvious to see that regardless of students' majors or school years, the levels of agreement on more state funding to soft matter research and the offer of college soft matter courses are identically above 6, which shows that most students agree that the soft matter education and research are important after watching the video. Among all the student majors and school years, students who pursue physical, chemical, and biological engineering majors, undergraduate third- and fourth-year students, and faculties have slightly higher levels of agreement compared to other groups. This can be explained since some of the properties of soft matter, such as Brownian motion, diffusion, self-assembly of DNA coated colloids, and the manufacture of microtubule-based active fluid and kinesin motor clusters are directly related to these fields of study and people who are more sophisticated in these fields tend to be inspired more about the soft matter physical phenomena and therefore the importance of education and research on soft matter related knowledge.

To further determine the appropriate levels of difficulty of soft matter courses, ***Figure 4.4.1*** shows the statistics on the preferences of the difficulty of soft matter courses responded by students with different majors **(C)** and school years **(D)**. It is obvious to see that for all students pursuing science and engineering majors, most of them agree that soft matter should be taught in an advanced course (3000 or 4000 level). However, students' preferences on soft matter courses difficulty differs from each range of school ages. From ***Figure 4.4.1*** **(D)**, most undergraduate students support that soft matter should be taught in an advanced course (3000 or 4000 level) whereas graduate students and faculties believe that soft matter should be taught in an intermediate course (2000 level). This can be explained by the fact that graduate students and faculties have learned more professional knowledge in their fields compare to undergraduates, which makes the introductory concepts of soft matter relatively easy to them.

To further understand how to distribute the soft matter related educational resources to different departments, ***Figure 4.4.2*** and ***Figure 4.4.3*** show the statistics on the preferences of the appropriate departments to offer soft matter courses responded by students with different majors and school years, respectively. We can see that regardless of student majors, more than 65 % of students agree that soft matter courses should be offered in physics, biomedical engineering, and chemistry and biochemistry departments, which are exactly the fields that involves the research on soft matter. The similar consensus holds for students in different school years except less percentage of graduate students think that soft matter courses should be offered in the chemistry and biochemistry department. It is also noted that the percentage of physical engineering students who believe mechanical engineering and mathematics departments should offer soft matter courses is considerably higher compared to other majors of studies.

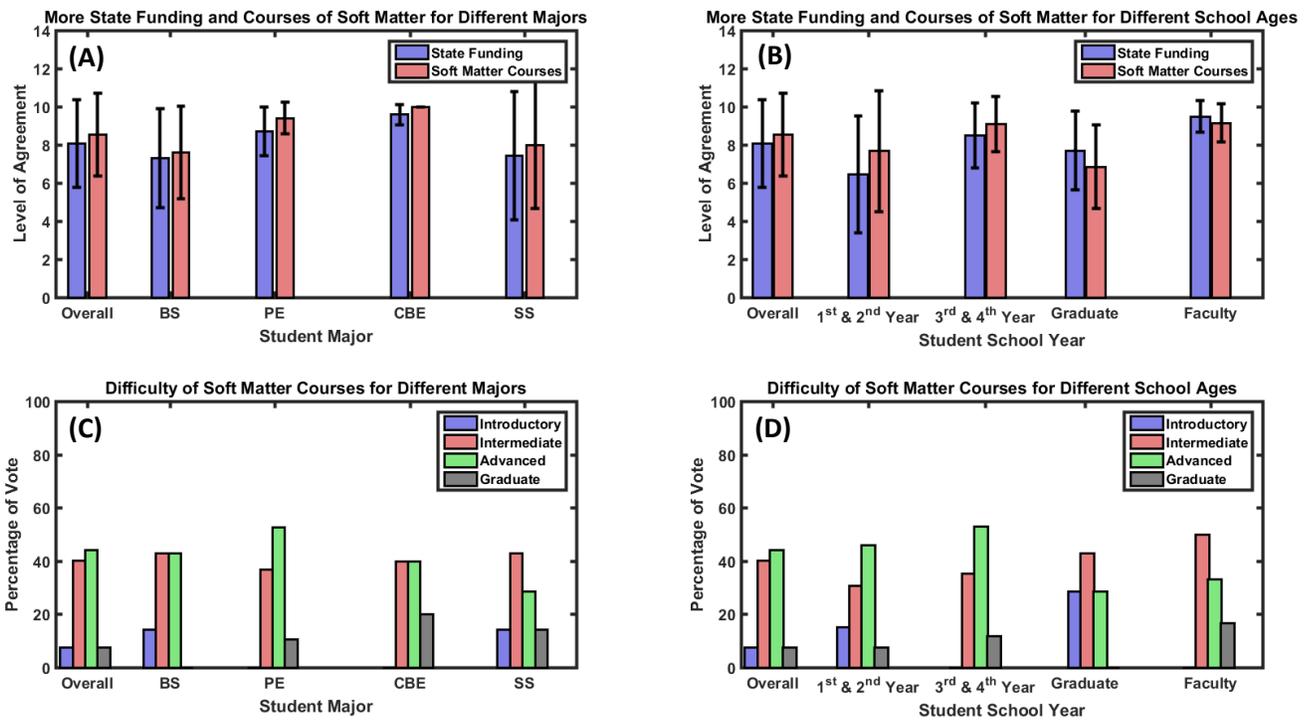

**Figure 4.4.1**: Levels of agreement on more government funding to soft matter research and the offer of college soft matter courses after watching the video responded by students with **(A)** different majors **(B)** school years; students' preferences on the difficulty of soft matter courses responded by students with **(C)** different majors **(D)** school years

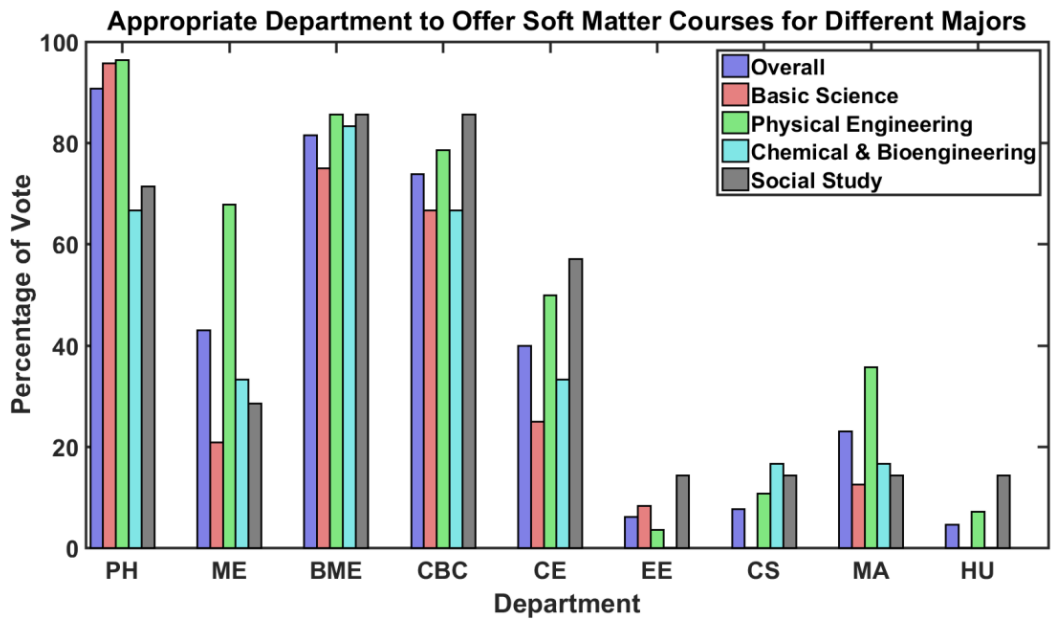

**Figure 4.4.2**: Statistics on students' preferences on the appropriate departments to offer soft matter courses responded by students with different majors

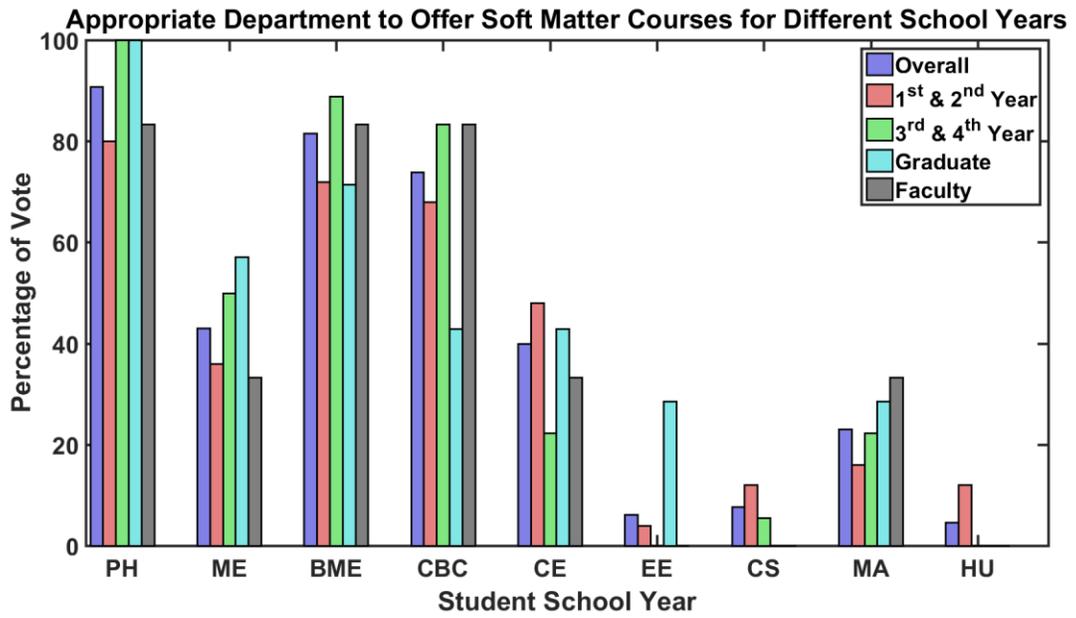

**Figure 4.4.3**: Statistics on students' preferences on the appropriate departments to offer soft matter courses responded by students in different school years

# Chapter 5: Conclusion

In this study, we composed a video to introduce soft matter to students, faculty and the general public, along with a survey to evaluate the effectiveness of the video on promoting the awareness of soft matter. The survey revealed that the video effectively increased the public interest in soft matter and motivated the participants to pursuit after the associated career, across disciplines, ages, and career goals. Moreover, the participants agreed with the need of more government investment and soft matter education, suggesting that WPI provide more resources on developing soft matter education, and that the leaders of WPI research groups provide more soft matter related research opportunity. However, this study only included 65 participants, due to the limitation of time in this project. A larger size of data will be required to consolidate the video assessment along with the survey outcomes. Our studies lay the groundwork for WPI to further promote soft matter research and education on its way to become a soft matter-vibrant center.